# Somatic Development Analysis of Handball Players using Computer Programs

**Assoc.Prof. Tiberiu Marius Karnyanszky, Ph.D.Eng.**
**Assist.Prof. Corina Musuroi, M.D.**
**Instructor Carla Amira Karnyanszky**
**"Tibiscus" University of Timişoara, România**

**ABSTRACT.** Assessing the degree of the child's body building optimal for obtaining performance in the practiced sport can be done by reporting the individual developing model to the typological models considered optimal for the requirements of each sportive branch in part.

## 1. Introduction

These characteristic typologies are defined by a series of criteria from which of a special importance are those correlated to the physical development. These are somatic-metric parameters of the body, specific to each individual, which depend to a large extend on the particularities hereditarily inherited and less on their modeling by training. The addressability to the performance sport is high during childhood, the parents being ready to invest in their children's practicing a sport in an organized manner, even if not at performance level. They have not always the initiative to do this or sometimes the chosen domain does not correspond to the child's or teenager's constitutional type.

These problems could be solved by assessing the children's physical development, in an organized manner, in schools, as a screening test. The results of the tests can make a data basis regarding the biological potential of the children population at the age for their selection for the performance sport. This data is useful both for the Physical Education teachers in schools, for the trainers at the sportive clubs and for the parents who are interested in their children physical activity.

The present study presents a testing model for the orientation towards the handball game using, consequently, the parameters of the optimal constitutional type for this game and the physical development criteria used in the selection for the performance handball.

**1. Objectives**

The assessment of the child's physical development presents a special practical importance in the Physical Education activity and sports. This assessment is the working instrument which can be used both in schools, for characterizing the children population development from a somatic point of view, and in the clubs for analyzing the sportspeople's performance potential.
Synthetically, the objectives of this analysis are:
1. *Establishing the physical development degree according to the child's age and sex:*
This activity has a practical component of the clinical, morphological and functional examination, and a theoretical component of analyzing and interpreting the results obtained by clinical observation.
- The general morphologic examination – which emphasizes the general harmony of the body development
- The anthropometric examination – consists of measuring the anthropometric parameters
- The biometric examination – consists of interpreting the results obtained by anthropometric determinations.

The theoretical part, the one of analysis by a comparative study can be done through:
- Comparison of the measured parameters to the average values of the referential population, values which are periodically established by the institutions specialized in medical statistics.
- Calculation of the proportionality indexes using the measured anthropometric parameters and comparing them to the normal values given by the specialty literature.
2. *Establishing the sportsperson's constitutional typology according to the practiced sports.*
The analysis is made by comparing the somatic development pattern of the assessed subject to the optimal constitutional type of the performance sportsperson for the considered sportive branch. This optimal pattern is established by the sportive medicine used in the selection for the performance sport.

The obtained data through these assessments are numerous, considering that it regards the entire children population, from the pre-university educational institutions. The processing and interpreting of the data supposes to pass through all the sages each time, meaning that a periodical examination becomes under these circumstances difficult. In order encounter these difficulties, the present study presents an operational pattern for a calculation program conceived to carry on the part of data analyzing and interpreting so that the entire procedure should become a routine activity in the school and sportive medical offices.

## 3. Material and method

The purpose of this study is to present a technical method based on using a calculation program easy to be used for analyzing the body development of the children and young persons, according to the age and sex, with the signalization of a favorable corporal construction for practicing a performance sport. .

For the efficient presentation of the method, the chosen subject was the seventeen years old student R.D., a performance sportsman in the handball team of the „Bega" School Sportive Club of Timişoara. The somatic-metric parameter were measured at the sports court as the sportive base has a well-equipped medical office

The following anthropometric parameters were followed::
1. *stature ( in cm)* – represents the measured distance from the  vertex  to the supporting plan of the inferior limbs;
2. *span (in cm)* – represents the distance between two digital point when the arms are in abduction (stretched out), with the elbow and the arm joints extended and the palms oriented to exterior so that the superior limbs segments should be on the same line.
3. *biachromial diameter (in cm)* – represents the distance between the two achromial points. The measurement is made on the individual in orthostatic position, keeping the superior limb relaxed along the body, by placing the two arms of the anthropometric compass at the level of the two achromial points.
4. *bitrochanterian diameter (in cm)* – represents the distance between the two trochanterian points. The determination is done on the

individual in orthostatic position, with the knees extended and the heel stuck together.

5. *the length of the palm (in cm)* – represents the distance between the styloid apophysis of the radius and the tip of the middle finger at the level of the arm;
6. *the length of the bust (in cm)* – is defined by the distance between the vertex to the sitting surface.
7. *the length of the superior limb (in cm)* – represents the distance between the achromial point and the digital point of the limb on the same side;
8. *the length of the inferior limbs (in cm)* – represents the distance from the antero-superior iliac spine to the ground, from which a correction index (35mm) is subtracted.
9. *the thorax perimeter (in cm)* - represents the length of the circumference of the thorax which varies according to the breathing phase adopted by the measured subject ;
10. *the perimeter of the limb's segment in cm (arm forearm, thigh, and shank)* – the circumference of each segment on a member is measured at the level of the most prominent muscular part of the region. Based on this values, the analysis of the somatic development was done, analysis which took two stages:

*The 1st Stage*: The qualitative and quantitative assessment of the corporal construction by reporting the values obtained in the study to those corresponding to the age and the sex in the referential population

*Quantitative assessment* – consists of comparing the values of the somatometric measured parameters with the average values in the referential population, according to the age and sex of the subjects in study.

The average values in the referential population of the somatometric parameters used in the evaluation of the children's corporal development are presented in table 1. In order to get them up-to-date, we considered the values presented by the „Victor Babes" Medicine and Pharmaceutics University of Timisoara, established in a PhD study for 2003.

If the measured parameters fall within the limits of +/- 2 DS comparing to the average ones the child enters the category of those normally developed for their biological age. If they go beyond those limits, as a general phenomenon, a deviation from the normal development is signalized; over +2DS indicates an accelerated growth and under -2DS, a physical development retard.

*Qualitative assessment* – consists of the comparison of the proportionality indexes calculated for the values of the measured parameters of the subjects in study; the values indicated by the specialty literature as optimal for a proportional development.

Table No.1 Somatometric parameters

| No. | Abbreviation | 17 years | | |
|-----|--------------|----------|------|-------|
|     |              | -2 DS    | VM   | +2 DS |
| 1.  | St           | 162,5    | 175  | 187,5 |
| 2   | IB           | 80,6     | 87,2 | 93,74 |
| 3.  | RightSLL     | 62,13    | 75,2 | 88,27 |
| 4.  | RightarmL    | 16,36    | 19,4 | 22,43 |
| 5.  | Left LIL     | 94,5     | 105,8| 117,06|
| 6.  | D. Biachr.   | 33,3     | 38,6 | 43,8  |
| 7.  | D. Bitroch.  | 26,5     | 29,6 | 32,63 |
| 8.  | Span         | 166      | 189  | 211,9 |
| 9.  | T Perim.     | 86,1     | 92,7 | 99,3  |
| 10. | Abd. Perim.  | 71,3     | 76,5 | 86,7  |

The indexes used in this study were:

- Giufrida Ruggeri Index: Bust / St x 100 (%)
- Adrian Ionescu Index: $I_{AI}$ = Bust – St / 2
- The report index of the SL to the stature: SLL / St x 100 (%)
- The report index of IL to Stature: LIL / St x 100 (%)
- The report index of the span to the stature: Span./ St x 100 (%)
- The report index of the biachrom. diam. to the stature:
  Biachr. D/ St x 100 (%)
- The report index of the bitroch.to the stature: Bitroch./St x 100 (%)
- Brugsch-Goldstein Index $I_{BG}$: ThPerim / St x 100 (%)
- The report index of the abd. perim.to the stature:
  Abd.Perim. ./ St x 100 (%)
- The report index of the arm perim.to the stature: A.Perim. / St x 100 (%)
- The report index of the forearm perim.to the stature:
  FaPerim. ./ St x 100 (%)
- The report index of the tight perim.to the stature: TgPerim./St x 100 (%)
- The report index of the shank perim. to the stature:
  ShPerim. / Stx 100 (%)

The average values of the proportionality anthropometric indexes for evaluating the physical development for the 16-18 years old period are presented in table 2:

Table No. 2 Average anthropometric values

| No. | Anthropometric index (%) | 16 Years | 17 years | 18 years |
|---|---|---|---|---|
| 1. | GR I: Bust / St x 100 | 51.60 | 51.67 | 51.74 |
| 2. | I. AI: $I_{AI}$ = BusSt – St / 2 | 2.78 | 2.93 | 3.08 |
| 3. | SLL / St x 100 | 42.84 | 42.68 | 42.79 |
| 4. | LIL / St x 100 | 48.39 | 48.32 | 48.25 |
| 5. | Span./ St x 100 | 101.57 | 101.24 | 101.16 |
| 6. | Biach.D/ St x 100 | 22.10 | 22.10 | 22.19 |
| 7. | Bitroch.D/St x 100 | 17.51 | 17.74 | 17.85 |
| 8. | BG I: Th Perim ./ St x 100 | 48.49 | 49.64 | 50.11 |
| 9. | Abd. Perim. ./ St x 100 | 39.66 | 39.77 | 40.49 |
| 10. | A. Perim. / St x 100 | 14.54 | 14.60 | 15.31 |
| 11. | Fa Perim./ St x 100 | 13.77 | 13.95 | 14.50 |
| 12. | Tg Perim../ St x 100 | 29.67 | 29.61 | 29.76 |
| 13. | Sh. Perim. / St x 100 | 20.04 | 19.98 | 20.11 |

The interpretation of the variations was the following:
- If the determined values fall within the referential values in the table, *with variations that do not overpass in plus or in minus a year* - the subject has a harmonious physical development suitable for his/her age.
- If the calculated values fall immediately on the next age year, with the same variation limits, the subject presents a *harmonious bur accelerated development.*
- If the calculated values fall on the one year inferior line, the subject has a harmonious yet deleted physical development.

*The 2nd stage:* The assessment of the degree in which the child's corporal construction is optimal for the performance handball..

This analysis was done by comparing the child's physical development pattern to the optimal somatic criteria established for the handball selection.

For the performance handball, there are considered the following *physical development* criteria (only part of them were used in the present test):
- Stature (St), over 190 cm, the subject enters this tendency considering the fact that, at his age, he has not achieved his maximum height.

- The span (S), 6 to 10 cm over the stature,
- Lumbar force – over 200% of his weight;
- The palm length - 26-28 cm; for his age over the average;
- Great aerobe force, (67-70 ml $O_2$/ kg body);
- Great anaerobe capacity (40-46 kgm/kg body);
- Palm length 1/8 - 1/9 of the stature;
- Very good neuro-motility coordination; the criterion is met.
- High skills and intelligence; the criterion is met;
- High speed, force, and resistance; the criterion is met.

## 4. Results

The measured anthropometric parameters for the 17 years old subject, male, are presented in table 3:

**The 1st stage – the qualitative and quantitative assessment of the physical education:**

In order to facilitate the input of the results of the proportionality indexes in the table, the following expression convention has been used:

- If the values obtained by calculation fall in the area of the lower ages than the subject's one, the term „under interval was used"
- If the values obtained by calculation correspond to the subject's age interval +/- 1 year, the term „within the interval" has been used;
- If the values obtained by calculation fall in the area of the higher ages than the subject's one, the term „over the limit" was used.

Table No. 3 Measured anthropometric values

| The name of the parameter | Value (cm) |
|---|---|
| Stature (St) | 186 |
| Weight (We) | 81 |
| The height of the bust (B) | 96 |
| The length of the superior limb (SLL) | 81 |
| The length of the inferior limb (LIL) | 108 |
| Span (S) | 195 |
| Biachromial diameter (Biacr.D) | 44 |
| Bitrochanterian diameter (Bitr.D) | 31 |
| Thorax perimeter (Thorax P) | 96 |
| Abdominal Perimeter (Abd.P | 80 |
| Arm Perimeter (AP.) | 38 |
| Forearm Perimeter (FaP) | 31 |
| Tight Perimeter (TP | 54 |
| Shank Perimeter (ShP) | 37 |

The comparison of the measured values to those from the literature is presented in table 4:

Table No. 4 Interpretation of values

| No | Abbre-viation | 17 years | | | Measured values | Interpretation |
|---|---|---|---|---|---|---|
| | | -2 DS | VM | +2 DS | | |
| 1. | St | 162,5 | 175 | 187,5 | 186 | Over average |
| 2 | IB | 80,6 | 87,2 | 93,74 | 96 | Over limit |
| 3. | RightSLL. | 62,13 | 75,2 | 88,27 | 81 | Over average |
| 4. | RightarmL | 16,36 | 19,4 | 22,43 | 26 | Significant over average |
| 5. | Left LIL | 94,5 | 105,8 | 117,06 | 108 | Over average s |
| 6. | BiAchr.D | 33,3 | 38,6 | 43,8 | 44 | Over average |
| 7. | Bitroch.D | 26,5 | 29,6 | 32,63 | 31 | Over average |
| 8. | Span. | 166 | 189 | 211,9 | 195 | Over average |
| 9. | T Perim.. | 86,1 | 92,7 | 99,3 | 96 | Over average |
| 10. | Abd Perim. | 71,3 | 76,5 | 86,7 | 80 | Over average |

The majority of the measured values of the anthropometric parameters are over the average, emphasizing a physical development motivated by the activity period as a performance sportsman. Some parameters as the span, the biachromial diameter or the length of the palm have values over the +2D indicating an accelerated increase of these dimensions. These parameters correspond to the assessment criteria for the performance in handball; therefore the sportsman potential for performance in this sport is confirmed.

The interpretation of the calculated proportionality indexes was done through a calculation program. According to table 4, we obtained results as presented in Table 5.

The length and the diameters of the limb, as well as the biachromial and bitrochanterian diameters, the span and the abdominal and thorax perimeters reported to the stature present greater values than the variation interval considered normal. The majority of the indexes are over the 17 years old column, meaning that the young man has a proportional, but accelerated growth of his trunk.

Table No. 5 Results and interpretation

| No | Anthropometric index (%) | 16 years | 17 years | 18 years | Calculated values | Interpretation of the values of the limit |
|---|---|---|---|---|---|---|
| 1. | GR. I: Bust / St x 100 | 51.60 | 51.67 | 51.74 | 51.61 | Within interval |
| 2. | AI.I : $I_{AI}$ = Bust – St / 2 | 2.78 | 2.93 | 3.08 | 3.00 | Within interval |
| 3. | SLL /St x 100 | 42.84 | 42.68 | 42.79 | 43.55 | Over |
| 4. | LIL /St x 100 | 48.39 | 48.32 | 48.25 | 49.46 | Over |
| 5. | Span./St x 100 | 101.57 | 101.24 | 101.16 | 104.84 | Over |
| 6. | Biachr D./St x 100 | 22.10 | 22.10 | 22.19 | 23.66 | Over |
| 7. | Bitroch.D/St x 100 | 17.51 | 17.74 | 17.85 | 18.28 | Over |
| 8. | BG.I: Thorax P./ St x 100 | 48.49 | 49.64 | 50.11 | 51.61 | Over |
| 9. | Abd. Perim./St x 100 | 39.66 | 39.77 | 40.49 | 43.01 | Over |
| 10. | A.Perim. /St x 100 | 14.54 | 14.60 | 15.31 | 20.43 | Over |
| 11. | FaPerim. /St x 100 | 13.77 | 13.95 | 14.50 | 16.67 | Over |
| 12. | T Perim./St x 100 | 29.67 | 29.61 | 29.76 | 29.03 | under |
| 13. | Sh.Perim. /St x 100 | 20.04 | 19.98 | 20.11 | 19.89 | Within interval |

**The 2nd stage – checking the criteria regarding the selection for the performance handball**

The results of the interpretation:
o The height, meaning the stature, (ST), over de 190 cm – the subjects presents this tendency, considering the fact that at his age he has not achieved the maximum;
o The span (S) with 6 to 10 cm over the stature – the subject has a plus of 9 cm
o The palm length 26-28 cm; by further measuring a length of 26 cm was obtained which is, for his age, over average.
o The thorax perimeter – 106; the subject has a circumference of 96, indicating an insufficient development;
o The arm perimeter (+1,0 cm) – it was assessed by the proportionality index which is over the variation interval;
o The tight perimeter (+1,7) - it was assessed by the proportionality index which is over the variation interval;

165

- o The shank perimeter (+1.9 cm) - it was assessed by the proportionality index which is over the variation interval;

There can be observed that the values of the somatic parameters mentioned in the performance handball criteria correspond to the selection ones.

## 5. Conclusions

The presented study aimed to describe the modality of using a technical instrument in the assessment of the children's and the youth's somatic development. The input data are measurements of the anthropometric parameters which are determined for each subject. The results obtained by analysis can be used both for assessing the growth and development of the children population, according to the age and sex, and for the children selection for the performance sports.

The analyzed subject is a 17 years boy, a handball player, at a performance level. The comparison of the measured values of the somatic parameters with the average ones emphasized the fact that the subject has a globally accelerated somatic development, with an accentuation of the phenomena for some dimensions.(the stature, the palm). The phenomenon is explainable by the fact that he is an active sportsman and the intensive physical effort determines the accelerated somatic development.

The analysis regarding the selection for practicing handball, which is positive in this case, is verified by the fact that he is a performance handball player.

Under these circumstances, considering the obtained results, it can be stated that the calculation program employed in this study can be used for a screening test on the children, being very useful to all the interested categories, especially to the Physical Education teachers and trainers, in their direct activity, in school or in the performance sports.